\documentclass[conference]{IEEEtran}

\usepackage{graphicx}
\usepackage{lineno,hyperref}
\usepackage{amsmath}
\usepackage{multirow}
\usepackage{subfigure}
\usepackage{xcolor}
\usepackage{comment}

\pdfoutput=1
\usepackage{ifpdf}
\usepackage{subfigure}
\usepackage{graphicx}
\DeclareGraphicsExtensions{.pdf,.jpg,.png}
\usepackage{amssymb}
\usepackage{array}
\usepackage{mdwmath}
\usepackage{mdwtab}
\usepackage{pslatex}
\usepackage{xcolor}
\usepackage{textcomp}
\usepackage{placeins}
\usepackage{eqparbox}
\usepackage{url}
\usepackage{stfloats}
\usepackage{multicol}
\usepackage{subfigure}
\usepackage{float}
\usepackage{algorithmicx}
\usepackage{algpseudocode}
\usepackage{amsthm}
\usepackage{comment}
\usepackage{todo}
\usepackage{color}
\usepackage{balance}
\usepackage{cite}
\usepackage[normalem]{ulem}

\usepackage{cite}
\usepackage{url}

\begin{document}
%
\title{IoT Technologies for Augmented Human: a Survey}

\author{\IEEEauthorblockN{Rustam Pirmagomedov}
\IEEEauthorblockA{Tampere University\\\
Tampere, Finland\\
Email: rustam.pirmagomedov@tuni.fi}
\and
\IEEEauthorblockN{Yevgeni Koucheryavy}
\IEEEauthorblockA{Tampere University\\\
	Tampere, Finland\\
	Email: evgeny.kucheryavy@tuni.fi}}


%


\maketitle

\begin{abstract}
Internet of Things (IoT) technology has delivered new enablers for improving human abilities. These enablers promise an enhanced quality of life and professional efficiency; however, the synthesis of IoT and human augmentation technologies has also extended IoT-related challenges far beyond the current scope. These potential challenges associated with IoT-empowered Augmented Human (AH) have so far not been well-investigated. Thus, this article attempts to introduce readers to AH concept as well as summarize notable research challenges raised by such systems, in order to facilitate reader’s further interest in this topic. The article considers emerging IoT applications for human augmentation, devices and design principles, connectivity demands, and security aspects.
\end{abstract}


%
\IEEEpeerreviewmaketitle

\section{Introduction}

Present efforts in human augmentation (sometimes referred to as “Human 2.0”) focus on the creation of cognitive and physical improvements as an integral part of the human body \footnote[1]{https://www.gartner.com/it-glossary/human-augmentation/}. These improvements are enabled by specially designed devices, such as leg or hand prosthesis, implants, artificial vision connected to the neural system of an organism, augmented reality glasses, hearing aids, and insulin pumps. Artificially recreated or extended abilities may improve quality of life and even give some competitive advantages for users.

Currently, the progress in human augmentation is driven by the interconnected Internet of Things (IoT) devices. The performance of these devices relies heavily on communication technologies. Commonly, such devices are located in close proximity to the human body. More specific applications may utilize bio-integrated devices, for example, neurally-controlled artificial limbs. All the devices used by an individual form an integrated ecosystem and should work coherently, which enabled by appropriate communication technologies. Depending on the type of the device the utilized communication technology may vary from traditional wireless protocols such as Bluetooth or Wi-Fi to highly specific technologies, such as electromagnetic or molecular nanonetworks. Therefore, a network of assisting devices can be considered as a highly heterogeneous Body Area Network (BAN). In addition to local communication, applications of Augmented Human (AH) require an internet connection (e.g., to be aware of context, offload of difficult computational tasks, upgrade software). As a whole, the concept of the Augmented Human creates a new segment of communication challenges, since the reliable performance of communication technologies in such systems is the essential enabler for the users' well-being.

Presently, the research on AH is spread across many different communities. From the perspective of communication technologies, devices for human augmentation have a lot in common with wearable electronics. However, being a branch of the IoT concept, AH devices perhaps provide the most critical class of services, because humans do not exist independently but rather as a part of human-centric AH systems in which one is trained \cite{xia2013design}. A failure in an AH application would cause chaos in this system and make the human vulnerable. Inherently, a communication failure will reduce a user’s physical or cognitive abilities. Thus, the AH applications require comprehensive analysis from the technological perspective to define their place in emerging network services. 

This paper provides an overview of the IoT technologies for AH and defines relevant research challenges in this innovative area.

The article is organized in the following way: Section \ref{Applications} provides an overview of AH applications. Section \ref{Design} considers the aspects and trends in device design. In Section \ref{Connectivity} we consider connectivity demands of the AH applications. Section \ref{Security} we discuss the security concerns for AH applications, and conclude the article in Section \ref{Conclusion}.

\section{Applications of AH Technologies} \label{Applications}

\begin{figure*}
	\includegraphics[width=\linewidth]{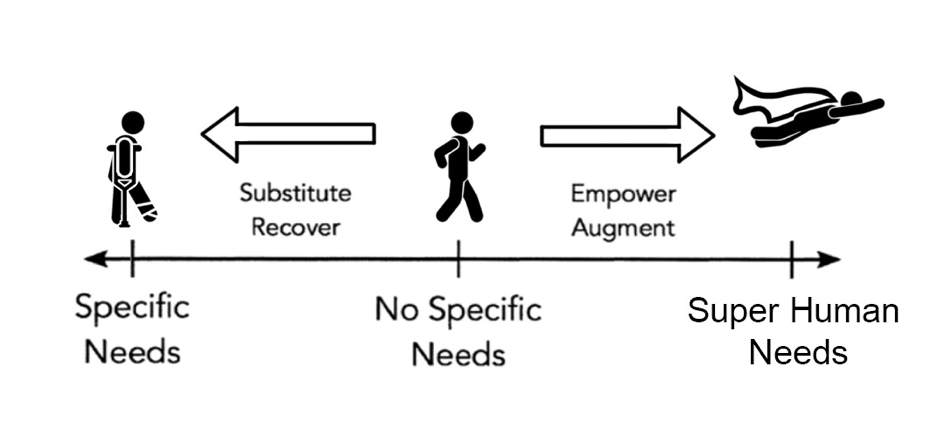}
	\caption{Assistive augmentation continuum according to \cite{nanayakkara2018augmented}}
	\label{continium}
\end{figure*}

Attempts to recover or improve human abilities began in ancient times. The majority of these attempts aimed at replacing a lost body part with an artificial one, for example, a leg or hand prosthesis. Some enthusiastic inventors aimed to go beyond the natural capabilities of the human organism by developing ``upgrades", such as wings for flying. These two vectors of augmentation development are still relevant and form an augmentation continuum as shown in the Fig.\ref{continium}. Initially, the majority of efforts towards human augmentation were focused on the improvement of physical abilities, while in the 20th century, due to progress in microelectronics, augmentation has been extended by advanced sensing and cognitive improvements. Small-sized electronic devices are capable of assisting in performing specific tasks, e.g., a hearing aid assists people with auditor disorders or the use of AR glasses capable of providing both navigation support and object recognition.

\subsection{Objectives of AH applications}
\begin{figure}[b]
	\includegraphics[width=0.95\linewidth]{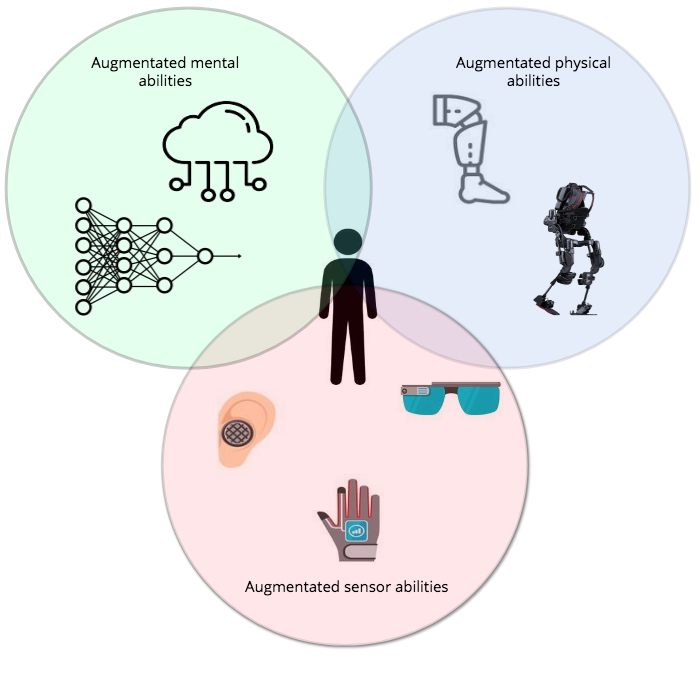}
	\caption{Areas of a human augmentation}
	\label{3areas}
\end{figure}

In a general case, AH systems assisting in daily routines \cite{maslow1943theory} using electronic devices, which are connected to a single BAN via different communication technologies. 
The network of interconnected wearables serves as a technological layout for high-level applications of AH, which enable physical augmentation, advanced sensing, and mental assistance Fig. \ref{3areas}. 

Physical augmentation aims at enhancement of an individual’s ability to move and manipulate objects. Examples of tools used for physical augmentation include exoskeleton, artificial arms and legs, or even a jet pack. Failures in physical augmentation can be hazardous for human safety and health; therefore physical augmentation tools should be capable of providing basic functionality even when network service or other resources are unavailable.

Sensory abilities allow a person to be aware of the environment and the context surrounding them. Sensory abilities may include vision, touch, hearing, smell, and taste. Augmentation may facilitate these senses by amplifying them or, in the case of having lost a sense, augmentation allows a transformation of the characteristics of one sensory modality into stimuli of another sensory modality \cite{leo2017computer}, e.g., visualizing speech or smells.  

Mental (or cognitive) augmentation provides data processing assistance and facilitates decision making. An illustrative example of a cognitive augmentation is a personal planning application, where users can save time and resources when planning daily routines. The application may plan optimal logistics during the day, select and book lunch at the highest quality and yet affordably priced restaurant within the defined location, find parking spaces and car charging plugs, integrate recommended physical activity into the day’s timeline, and automatically revise plans in accordance with changing conditions (automatic negotiations with involved parties and reconfiguration of schedule). All these functions can be performed in a background mode, increasing the efficiency of a working day and saving time for creative activities or leisure. Currently, cognitive augmentation is the most familiar branch of human augmentation because of its widespread use in mobile applications. As one may observe, technologically such applications rely on machine learning \cite{sekaran2019improving} and entirely hinge on information about the environment, while also being significantly dependent on an internet connection.

\subsection{Classification of AH applications}

Taxonomy of AH applications include three major classes: (i) supporting independent living (e.g., for the aging population or people with impairments); (ii) facilitating the professional performance; (iii) self-efficiency and entertainment.

The applications which support independent living allow users to satisfy their basic daily needs without the assistance of other people. In addition, such applications monitor users' health conditions in real-time and increase their safety (e.g., by protecting aging people from occasional fall). As a result, nursing costs can be considerably reduced while also improving the quality of life for both aging and disabled people.

The AH applications for improving professional performance focus on augmenting the abilities relevant to the professional areas of an individual. For instance, an exoskeleton for a worker allows for moving heavy weights without harmful consequences for the spine. Another illustrative example comes from emergency response, where AH may enhance the performance of rescue team members by providing augmented sensing (e.g., sensing of hazard gases, utilizing thermal vision), empowered physical abilities (e.g., exoskeleton), and efficient decision making (e.g., AI-assisted operation).

The entertainment class aims to provide unusual user experiences (e.g., flying with a jet pack) or an immersive experience of extreme situations (e.g., virtual reality gaming) without physical risks to the user.

\begin{figure}
	\includegraphics[width=1\linewidth]{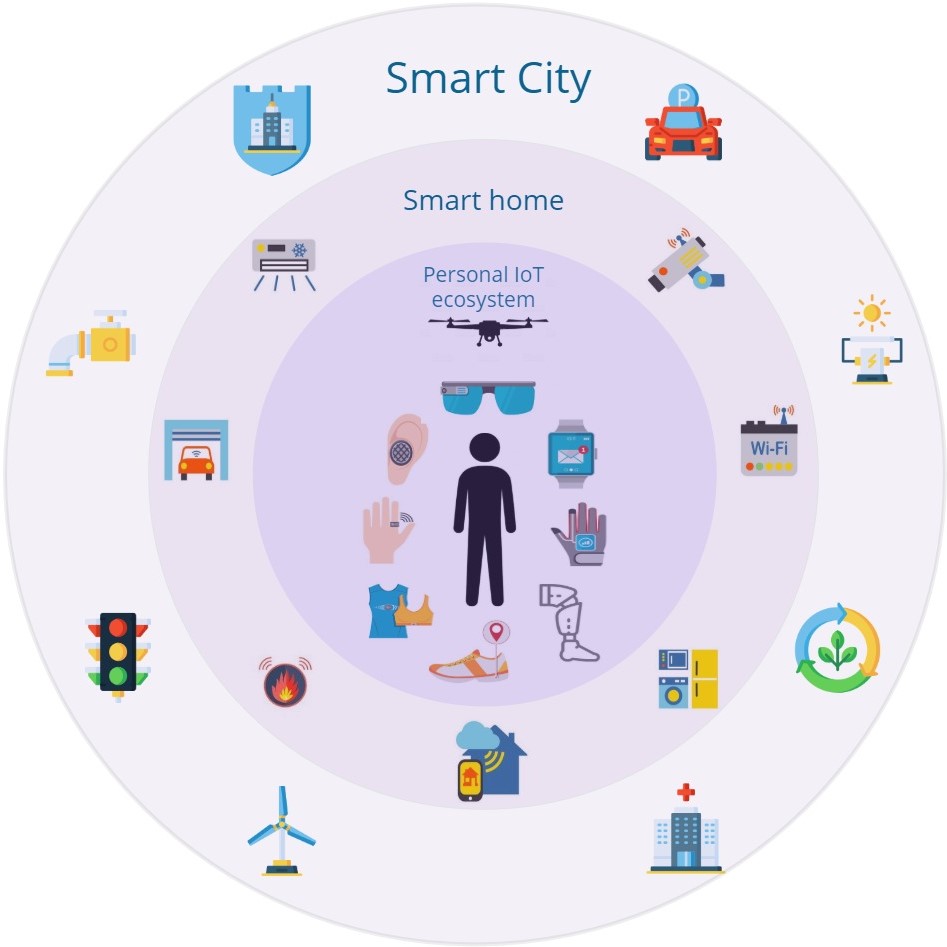}
	\caption{Integrated smart environment}
	\label{PHC}
\end{figure}

It is worth noting that AH systems may encompass the entire context where a person exists \cite{blackman2016ambient}, which include interaction with proximate entities such as buildings, city infrastructure, and other individuals. Communicating with each other, these form an integrated smart environment (Fig.\ref{PHC}).

\subsection{Challenges}

\textbf{Ethical and social aspects.} Innovative technologies for human augmentation will undoubtedly bring new challenges in ethical and social fields. Augmented features may become trendy especially among a younger generation\cite{koucheryavy2017quo}, or the border between artificial and natural abilities may become blurred. It is not evident which specific issues will become a part of the agenda, nevertheless, ethical and social aspects require primary consideration.

\textbf{Human-machine interaction.} User experience issues are of the utmost importance when designing IoT systems for human augmentation. The IoT-empowered AH systems should provide functional but straightforward user interfaces to avoid certain user groups (e.g., older people) from feeling uncomfortable when using such systems (due to the system complexity).

\section{Devices and Design Principles} \label{Design}

Technological challenges related to AH devices are primarily shaped by design principles, which rely on users' demands and expectations. Regarding the wearable IoT devices (which AH systems are), the users' expectations are primary: small size and weight of devices, enhanced reliability, and long battery lifetime. 

\subsection{Flexible Hybrid Electronics}
Flexible Hybrid Electronics (FHE) integrates devices from thinned flexible materials with electric circuits in formats that can be thin, light-weight, flexible, bendable, conformal, potentially stretchable and disposable \cite{lombardi2018copper}.  FHE offer notable advantages over the conventional electronic systems that are made of bulky and rigid materials \cite{herbert2018soft}. Recent advancements in the advanced materials and soft mechanics have enabled a successful integration of rigid, miniaturized chips with flexible/stretchable circuit interconnects. Such FHE in AH applications enhances signal processing, memory, and wireless power transfer in wearable systems \cite{valentine2017hybrid,roshan2017design}. For example, in real-time monitoring of health parameters, FHE enables bio-friendly devices on biological tissues, such as artificial human skin, or internal organs with time-dynamic motions \cite{gutbrod2014patient}. In general, implementation of FHE in AH enabling improved wearability and performance for the devices, and as a result, facilitating their use among individuals. 

\subsection{Reduced size of the devices}
Due to the progress in nanotechnologies, IoT wearables can be deployed at the nano level (named as Nanonetworks)~\cite{akyildiz2008nanonetworks}. Such nanodevices employ unique properties of graphene, which allows a significant size decrease for electronic elements, including antennas, processors, receivers and transmitters~\cite{jornet2013graphene,jornet2014graphene,moon2015zero}, as well as sensors and actuators~\cite{le2012graphene,chen2012enhanced}. The graphene-based nanoantennas enable communication in the THz frequency band~\cite{jornet2014graphene}. However, the distance of communication in the THz frequency band is substantially limited by the high signal power losses during propagation~\cite{akyildiz2010electromagnetic,kokkoniemi2015frequency}. The distance of communication will not exceed 2 meters, even in an air environment with minimal humidity; if the communication is performed in an environment with a high concentration of liquids, such as a human body, the distance of transmission will decrease to several millimeters~\cite{yang2015numerical} establishing new challenges related to enabling communication within such networks. These graphene-based devices (antennas and transceivers for THz communication) are small enough to be integrated into biological systems (on the border between the organism and the environment) and can be easily integrated into modern communication devices (e.g., smartphones) as they are based on existing electronic technologies.

\subsection{Improving energy effeciency}
A power unit used in wearables typically the most significant contributor to both the size and weight of the devices \cite{mosenia2017wearable}. As a consequence, developers must balance between size and the capability for autonomous operation when designing wearables. A majority of devices are designed with the priority given to size and weight, and thus have minimal operation time between recharging \cite{rault2014energy}. However, users' expectations continue to move toward fully autonomous devices without recharges or other maintenance operations. To address these demands, recent research efforts have targeted enhanced battery lifetime through improving the energy efficiency of the devices. Significant energy costs in wearables come from network functions, data acquiring, and processing \cite{nakhkash2019analysis,williamson2015data,javed2019taeo}.

The networking overheads of wearable devices was investigated in \cite{pirmagomedov2016simulation,behera2019residual}. More specifically, these works considered digital traffic generated by the wearable network in real-time mode. Results of the study demonstrated that network resource utilization in wearable systems is extremely low due to signaling overheads. However, the efficiency can be improved if an advanced data management algorithm is utilized on the BAN gateway. One such algorithm was proposed and evaluated in \cite{pirmagomedov2017dynamic}. The reported results demonstrated improved networking efficiency by approximately 80 percent via the reduction in network signaling overheads, while the performance of applications decreased negligibly. Despite the notable improvements in networking, the energy efficiency of the considered systems is far from optimal and has massive potential for further improvement.

From the perspective of data processing, a drastic improvement is the promise of Approximate Computing (AC) \cite{xu2015approximate}. Approximate Computing is inspired by the Pareto Principle according to which, roughly 80 percent of the effects come from 20 percent of the causes. Regarding the wearable networks, this principle can be formulated in the following way: capturing just 20 percent of the data may enable 80 percent of the application’s performance. It should be noted that the actual percentage can be different; however, the general principle remains the same – a minority of efforts provides the majority of results.

Wearable applications work with noisy data; thus they are natively resilient to error \cite{pirmagomedov2018iot}, moreover, most of the applications do not require extremely precise results, thus the paradigm of an acceptable margin of error as introduced by AC promises significant energy-efficiency gains for AH systems.

\subsection{Reliability}

Reliability issues need to be addressed long before a device could be considered for any mission-critical application. However, the reliability and validity of existing wearable devices is concerning. The majority of available devices are not verified in terms of accuracy and reliability \cite{piwek2016rise}. Recent tests among wearables showed significant variations of accuracy with error margins of up to 25 percent \cite{lee2014track}.

In addition to device reliability, but by no means less important, is the enabling quality of server platforms. Possible adverse effects from a cloud server failure are widely discussed in the literature \cite{cerin2013downtime,vishwanath2010characterizing,mosenia2017wearable}, and can be considerably mitigated via placement optimization \cite{zhou2016cloud}.

\subsection{Challenges}

\textbf{Developing networks of nanodevices.} Recent developments in nanotechnologies have enabled tiny-sized devices with both sensor and actuator functionality. However, due to multiple limitations, these devices are not capable of supporting standard communication protocols, including medium access control, routing, and security. Although networking among nanodevices is widely discussed in the literature, commercially available solutions have yet to be delivered, which keeps the door open for transferring theoretical findings to the real world.

\textbf{Power supply.} Emerging AH systems should fully utilize the benefits of efficient wireless power harvesting and energy transmission \cite{singanamalla2019reliable}, as well as low energy technologies, for reducing a user’s routine in its relation to the charging of devices. 

\textbf{Requirements for the devices and testing specifications.} Despite, the notable progress in provisioning reliable AH operation, there is a lack of systematic perspective on the reliability of mission-critical IoT systems. This gap is expected to be fulfilled by the efforts of international standardization bodies (e.g., SG11 ITU-T) which perform extensive work towards the standardization of unified testing procedures for such systems.

\textbf{Balancing the trade-offs between energy efficiency and accuracy.}
Implementation of AC promises a reduction of energy consumption by computing and sensing blocks of AH systems. However, the balance between energy efficiency and application performance should be clearly defined.

\section{Connectivity Demands of  AH} \label{Connectivity}

The connectivity demands of AH include intra-BAN and inter-BAN considerations and cover physical interfaces, networking architecture, and AH integration in emerging network infrastructure (5G/5G+).

\subsection{Multi-tier networking architecture}

\begin{figure}[h]
	\includegraphics[width=0.9\linewidth]{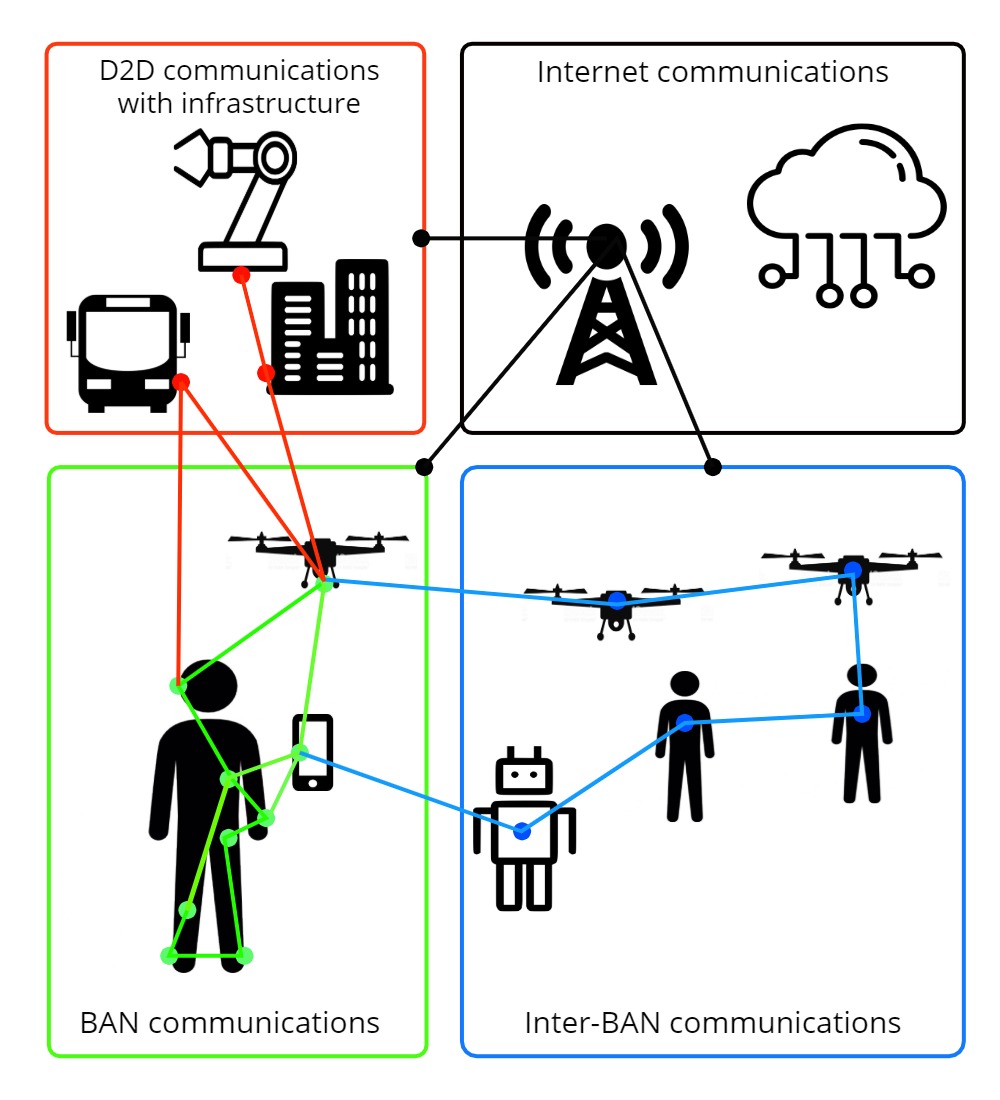}
	\caption{Integrated smart environments}
	\label{Multi}
\end{figure}

To enable the sustainable operation of AH devices, and context-awareness, AH systems must support multi-connectivity when operating in a multi-tier network environment (Fig.\ref{Multi}). 

Intra-BAN communications integrate all personal devices of an individual into one network. Such a network can operate in a distributed way or can be orchestrated by the head node (e.g., smartphone or body gateway). The orchestrated BAN is less reliable, because the fault of orchestrating devices causes a disruption of the whole BAN, while in a case of a distributed approach the network is resilient to the faults of individual devices~\cite{khan2012stitching}. On the contrary, an orchestrated BAN demonstrates better quality of service (QoS) and energy efficiency \cite{rashidi2012survey,ylisaukko2004five,poon2006novel}.

Inter-BAN communication covers the interaction between devices of two or more individuals. Such interaction often relies on device-to-device communication (D2D) and is required for enabling synchronization among AH systems when users collaborate. This type of communication is commonly characterized by a higher temporal and spatial dynamic (e.g., link blockages and outages). To improve the stability of sessions, the communication links can be established via assisting robot relays, such as drones. Drones can be considered as a part of a personal IoT ecosystem where they contribute to sensor augmentation, providing additional information about the environment. Simultaneously they may serve as relays for reliable D2D communication (e.g., connection with a user around the corner). In total, communication with infrastructure in direct mode allows an AH to be context-aware without a load on the mobile infrastructure. An internet connection via the mobile network infrastructure can be used for accessing cloud servers and other devices, which are not reachable via D2D communication (e.g. offloading of computation to the edge) \cite{kovtunenko2019multi}.

\subsection{Wired and wireless}

Wireless interfaces allow the creation of flexible connectivity within BAN. The tree of IEEE 802.15 standards specified wireless technologies adjusted for BAN use cases. Commonly used wireless technologies for intra-BAN communication include Bluetooth (IEEE 802.15.1), ZigBee (IEEE 802.15.4) and more recent WiMedia (IEEE 802.15.3) for ultra-wideband links \cite{mosenia2017wearable,singanamalla2019reliable}. Inter-BAN connections are enabled by Wi-Fi (IEEE 802.11) and mobile networks (e.g., LTE, 5G NR).

Wired interfaces enable improved reliability and stable quality of connections, which can be fundamentally important for critical elements of AH. Moreover, wired devices (almost unsusceptible to radio interference) reduce the problem of the radio-noisy environment when plenty of wireless devices are working in close proximity (ultra-dense scenario). In addition, a wired connection can be used for an energy supply which is a notable advantage of such systems. However, the low flexibility of wired networks significantly limits their utilization in AH systems. Wired connections are currently selected exclusively for intra-BAN communication. The most suitable niche for wired communications is related to cases where connected devices are not expected to be moved considerably in relation to each other. For example, elements of an exoskeleton, elements of smart textile, and sensors embedded in the skin and connected via smart tattoo.

Recent advances in inductive links and intrabody links may establish a new branch of communication technologies for AH systems, based on using human body tissues as a transmission medium (e.g., molecular communication) \cite{akyildiz2008nanonetworks,atakan2012body}.

\subsection{Increasing throughput}

Initially, wireless technologies for machine type communications (interconnected devices) were developing with a focus to low rate traffic (e.g., telemetry), and a limited density of devices in a network. Presently, due to the reduced size of wearables, the density of connected devices can be significant. In addition, their services have spread far beyond simple telemetry, and they now use media extensively (e.g., AR/VR video services). As a result, IoT devices are generating a notable portion of data in the network, which can be expected to continue to increase in the future.

Supporting a high data rate among wireless wearable devices, especially in dense deployment (e.g., crowded streets of the city, stadiums) is a challenging task. The primary concern is interference when many devices operate simultaneously. As an alternative to an extensively employed microwave spectrum, it is proposed to use millimeter-wave (mmWave) links \cite{venugopal2015analysis}. Due to the higher spectrum and less interference (because of greater signal loss at these frequencies), mmWave links are considered as a solution for the mitigation of interference and throughput concerns in emerging wearable networks  \cite{venugopal2016millimeter}.

\subsection{Augmented Human in 5G/5G+ landscape}

The connectivity challenges of AH in 5G/5G+ networks are driven by the spontaneous forming, maintaining and termination of heterogeneous networks of AH devices, and traffic flow balancing. Mission-critical communication has already been deeply investigated and discussed in the literature. However, network demands in the considered scenarios all occur in predefined locations (e.g., manufacturing, transportation hubs, medical facilities) \cite{olasupo2017link,brahmi2015deployment,orsino2017effects}, while network demands of AH applications are characterized by a high degree of temporal and spatial variations \cite{wang2017efficient}. Therefore, conventional “static” network planning methods are inefficient for AH and require development of adaptive methods.

In comparison with legacy mobile networks, 5G systems bring a considerable shift in the quality of services offering Ultra-Reliable Low-Latency Communication (URLLC) for delay-sensitive applications, which open new horizons for AH applications. More specifically, the dynamic demands of AH are expected to be addressed in 5G by utilizing mobile access points (e.g., cell on wheels, aerial access point) and traffic offloading on D2D mesh networks. Additionally, connectivity of AH can be considerably enhanced by utilizing multiband access (e.g., using sub-6 GHz and millimeter-wave bands of 5G NR simultaneously).  

Nevertheless, the mission-critical services natively supported by 5G systems require standardization efforts to meet AH demands. These efforts should result in a prioritized network service of AH applications and support interoperability between AH systems in 5G and beyond.

\subsection{Challenges}

\textbf{D2D mesh networking.} Secure inter-BAN multi-hop D2D communications are required for supporting merging and consequent splitting of AH systems employed by different users during their collaborative activities. Merging of AH systems means the incorporation of corresponding BANs. Such connectivity on the fly, requires robust devices identification method, neighbor’s discovery, routing, automatic choice and assignment of devices acting as a heterogeneous gateway to connect devices with different radio access technologies. For the last, but not least, it is essential to incentivize users to share their resources and participate in mesh networks; otherwise, the performance of the meshes will be very limited.

\textbf{Health Concerns.} Wide use of wireless wearables raises concerns related to the effects caused by the high-frequency electromagnetic waves on people’s health. The sensitivity of human tissues and skin to electromagnetic radiation, as well as long term effects caused by wireless devices, needs to be analyzed carefully.

\textbf{Enabling directional wireless communications in BAN} Directional wireless communication is extensively discussed in the literature as a feature of emerging air interfaces operating using high frequencies (e.g., millimeter-wave or THz communication). Utilization of directional antennas in wearable networks significantly increases the complexity of wireless interfaces, but promising lower interference among devices and gigabits-per-second rates (if mmWave links used)\cite{venugopal2016millimeter}. To enable directional links in BAN, research challenges related to beamforming techniques must be addressed \cite{galinina2016assessing}.

\textbf{Adaptive network management mechanisms.} A novel signaling architecture is required for capturing and predicting AH demands, in order to enable real-time network adaptation to varying demands of AH application and the varying available network resource. The promising solutions for addressing this challenge may come from the synthesis of machine learning approach and SDN/NFV technologies.    

\section{Security Considerations} \label{Security}

\begin{figure}[h]
	\includegraphics[width=0.9\linewidth]{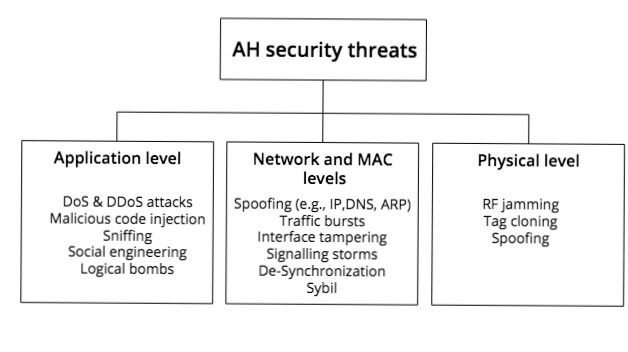}
	\caption{Common security threats in layered approach}
	\label{threats}
\end{figure}

Applications of AH bring security concerns to the top, as security breaches in such enablers can have dramatic results to both the infrastructure and the individuals who rely on them. International standardization bodies are considering security challenges architecturally \cite{recommendation2002security,zhang2017overview,jover2019security}. Following this, Fig. \ref{threats} summarized in a layered manner, common security threats relevant for AH applications.

\subsection{Physical level}

Attacks on the physical level may disrupt the normal operation of connected devices even if the high level (MAC, network and application) is well designed. For example radio frequency (RF) jamming may imply interruption of wireless communication using high power radio signals of the same frequency as used by AH devices. RF jamming may entirely block communication or interfere with it. The later may exhaust batteries of wearables due to additional energy costs required for numerous retransmissions, using higher transmit power, idle listening, etc. 

A wireless medium is essentially a broadcast one, which makes such systems vulnerable to eavesdropping (e.g., attackers may eavesdrop on ongoing transmissions and hijack the contents or spoof the other user) \cite{hussain2017security}. 

\subsection{Network and MAC levels}

A significant issue of network and MAC security is caused by a lack of robust device identification methods \cite{paranjothi2017survey,7581523,hussain2017security}. There are many solutions proposed for the identification \cite{samaila2018security}. These technologies can be classified into two groups: virtual identifiers and physical identifiers. Currently, the most popular identifiers (IMEI, MAC-address) are recorded into the memory of the device \cite{vladimirov2019unique}, which makes them vulnerable to cloning and tampering~\cite{hackIMEI, DefendIMEI, changemac, cherchali2015technique}. An alternative recently was proposed a concept of hybrid identifier \cite{vladimirov2019unique}, which is significantly more resilient to tampering and potentially may address the issue of reliable device identification in the network. A reliable identification method is required to enable the blocking of untrusted devices on MAC and network levels which reduces the risk of attacks based on accessing the network, including Sybil, tampering server or client interface, spoofing (e.g., DNS, ARP), signaling storms (redundant signaling messages), traffic bursts (e.g., extensive request or data forwarding), and de-synchronization. 

Concerning the network level, IEEE 802.15.6 defines three levels of security with the focus on critical applications of BAN. According to the standard, each security level has different properties, security levels, and data frame formats. The lowest level of security is provided on level 0, which employs an unsecured data frame for communication. This level has no mechanism for data integrity, confidentiality and privacy protection, and replay defense. The next level provides authentication for enhancing security; however, data is not encrypted. Thus confidentiality and privacy issues are not addressed. Finally, the third level enables authentication and encryption, providing maximal security. The required security level can be selected when a new device associates BAN. The security mechanisms proposed in IEEE 802.15.6 support both unicast and multicast \cite{kwak2010overview}.

\subsection{Application level}

Software (including firmware) quality and immutability are primary concerns at the application level. Most common attacks exploit the vulnerabilities of software to inject malicious code or logical bombs, performing DoS attacks, and sniffing. It is worth noting that untrusted software producers may incorporate malicious code or a logical bomb into their application by default, which can make the user vulnerable. Beyond this, the most successful attacks at the application level are based on social engineering. This type of attacks exploits users' weaknesses, which is often much easier than hacking a well-designed application.

\subsection{Challenges}

\textbf{Software secured from social engineering attacks.} Software utilized in AH applications should be design in such a way to a users' actions by enabling protection from social engineering attacks, by limiting their rights in a system. Recent machine learning algorithms are expected to enable monitoring and dynamic protection from social engineering attacks \cite{heartfield2016taxonomy}. 

\textbf{Standardization of requirements and testing procedures.} Security and reliability requirements for AH applications have to be standardized to provide a validated design framework for developers. New applications can then be considered as ready for AH services if an appropriate testing campaign has certified their conformity to the standard.    

\textbf{Device identification and validation} Counterfeit devices still have a notable share in the market. Such devices may operate incorrectly and reduce the performance of the system as a whole. Thus, it is especially important to provide a robust device identification system for AH applications. Such a system can be used for blocking counterfeit and untrusted devices in the network which facilitates the security of the applications.

\section{Conclusion} \label{Conclusion}

Communication technologies of the past decades notably shaped social and lifestyle changes. Internet-related technologies accelerated lifestyle via efficient and prompt information exchange. Currently, in the era of IoT one may observe how connected devices have become fully autonomous, delivering advanced services to their users. Emerging IoT applications are facilitating human augmentation via enhanced sensing, increased physical power, or cognitive performance. These applications form a new area for research and development, promising to become one of the most impactful technologies in the foreseeable future.

This paper covered the main aspects of IoT technologies for human augmentation and identified possible future research directions. The topic of human augmentation is highly interdisciplinary; thus, the defined challenges are not limited to communication technologies only, and their mitigation requires efforts in ethics, security, and natural sciences. Only collaborative work on this topic enables real opportunities for human wellbeing via IoT augmentation.

\bibliographystyle{IEEEtran}
\bibliography{mybibfile}

\end{document}